# Dust/ice mixing in cold regions and solid-state water in the diffuse interstellar medium

Alexey Potapov[1], Jeroen Bouwman[2], Cornelia Jäger[1], and Thomas Henning[2]

[1]*Laboratory Astrophysics Group of the Max Planck Institute for Astronomy at the Friedrich Schiller University Jena, Institute of Solid State Physics, Helmholtzweg 3, 07743 Jena, Germany, email: alexey.potapov@uni-jena.de*
[2]*Max Planck Institute for Astronomy, Königstuhl 17, D-69117 Heidelberg, Germany*

**Whether ice in cold cosmic environments is physically separated from the silicate dust or mixed with individual silicate moieties is not known. However, different grain models give very different compositions and temperatures of grains. The aim of the present study is a comparison of the mid-IR spectra of laboratory silicate-grains/water-ice mixtures with astronomical observations to evaluate the presence of dust/ice mixtures in interstellar and circumstellar environments. The laboratory data can explain the observations assuming reasonable mass-averaged temperatures for the protostellar envelopes and protoplanetary disks demonstrating that a substantial fraction of water ice may be mixed with silicate grains. Based on the combination of laboratory data and infrared observations, we provide evidence of the presence of solid-state water in the diffuse interstellar medium. Our results have implications for future laboratory studies trying to investigate cosmic dust grain analogues and for future observations trying to identify the structure, composition, and temperature of grains in different astrophysical environments.**

Cosmic dust grains with carbonaceous and siliceous composition [1,2] represent the most pristine starting material for planetary systems, influence the thermodynamic properties of the medium by absorption and emission of stellar light, and provide a surface for key astrochemical reactions. Dust grains in cold dense astrophysical environments, such as dense interstellar clouds, protostellar envelopes and planet-forming disks beyond the snow line, are typically considered to be mixtures of dust particles with molecular ices. Water is the main constituent of these ices accounting for more than 60% of the ice in most lines of sight [3]. Ices are believed either to cover the surface of a dust core and/or to be physically mixed with dust. While the first case, ice-on-dust, has been intensively studied in the laboratory in recent decades, the second case, ice-mixed-with-dust, presents practically uncharted territory.

The ice-on-dust case is typically modelled by ice mixtures deposited onto standard laboratory substrates, such as gold, copper or KBr, which are not characteristic of cosmic dust grains. In addition to these studies, there are studies on physical-chemical processes, such as formation, desorption, and diffusion of molecules, on the surface of cosmic dust analogues: amorphous carbon grains, atomic carbon foils, graphite, amorphous silica, and amorphous and crystalline silicate grains. The reader can find examples of and references to these experimental studies in recent papers [4-9].

Concerning ice-mixed-with-dust, this is a new direction of research started very recently by us [10-12]. One of the conclusions of our previous study of the optical constants of dust/ice mixtures [11] was that differences between measured constants and constants calculated using effective medium approaches show that a mathematical mixing (averaging) of the optical constants of water ice and silicates for the determination of the optical properties of silicate/ice mixtures can lead to incorrect results. Thus, in the case of dust/ice mixing in cold astrophysical environments, to reproduce observational spectra more reliable, one has to use constants (or spectra) measured for physical mixtures of dust and ice.

Evidence to date shows that ice is mixed with dust in space. The examination of the samples returned by the Stardust mission showed that the dust and water-ice agglomerates were mixed before cometesimals formed in the outer solar system [13]. Results of the Rosetta mission to the comet 67P/Churyumov–Gerasimenko demonstrated that the core consists of a mixture of ices, Fe-sulfides, silicates, and hydrocarbons [14]. Different models of comets, such as dirty snowball, icy glue, and fractal aggregates [15] mean that ice is mixed with dust in cometary nuclei. Asteroids are expected to retain ices in their interior [16]. Dust grains in astrophysical environments form fractal aggregates characterized by a very high porosity that is suggested by the analysis of cometary dust particles [17-20], dust evolution models [21-23], laboratory experiments on the dust particle aggregation by collisions [24,25], and on the gas-phase condensation of grains with their subsequent aggregation on a substrate [26,27]. It is reasonable to assume that ice fills the pores of grain aggregates. If grains in cold astrophysical environments are physically mixed with ices, in laboratory experiments, they should be better modelled by ice mixed with dust rather than ice layered on dust as typically assumed.

In view of new astronomical instruments, such as James Webb Space Telescope (JWST) and Extremely Large Telescope (ELT), which are expected to bring novel and detailed information on cosmic dust grains, there is a strong need for reliable laboratory data based on analogues of cosmic grains. This provides the motivation for studying the physical-chemical and spectral properties of dust/ice mixtures. The aim of the present study was a comparison of the mid-IR spectra of silicate-grains/water-ice mixtures with astronomical observations to evaluate the presence of dust/ice mixtures in interstellar and circumstellar environments.

**Results**

**Laboratory measurements on silicate-grains/water-ice mixtures**

Figure 1 shows IR spectra of the $H_2O$ and the $MgSiO_3/H_2O$ samples at 10, 50, 100, 150, and 200 K typical for all silicate-grains/water-ice samples studied, namely $MgSiO_3/H_2O$, $Mg_2SiO_4/H_2O$, and $MgFeSiO_4/H_2O$. The observed vibrational bands in the 10 K spectra are as follows: 3280 cm$^{-1}$ (3.05 μm) involving symmetric and asymmetric $H_2O$ stretching, 2200 cm$^{-1}$ (4.55 μm) − $H_2O$ combination mode, 1640 cm$^{-1}$ (6.10 μm) − $H_2O$ bending, 1040 cm$^{-1}$ (9.62 μm) − Si–O stretching, 770 cm$^{-1}$ (12.99 μm) − $H_2O$ librational motion, and 510 cm$^{-1}$ (19.61 μm) − O–Si–O bending (not shown).

The temperature increase leads to the structural transformation of water ice from high- to low-density amorphous ice in the temperature range between 10 and 80 K and from amorphous to crystalline ice around 140-150 K. While the 10 K spectra of the $H_2O$ and silicate/$H_2O$ samples are practically identical (except the Si-O stretching region), the temperature rise leads to an appearance of a high-frequency shoulder in the $H_2O$ stretching band and to only a very slight broadening of the $H_2O$ bending band for the silicate/$H_2O$ spectra compared to the $H_2O$ spectra. These differences appear after heating the samples and become more noticeable with the temperature increase.

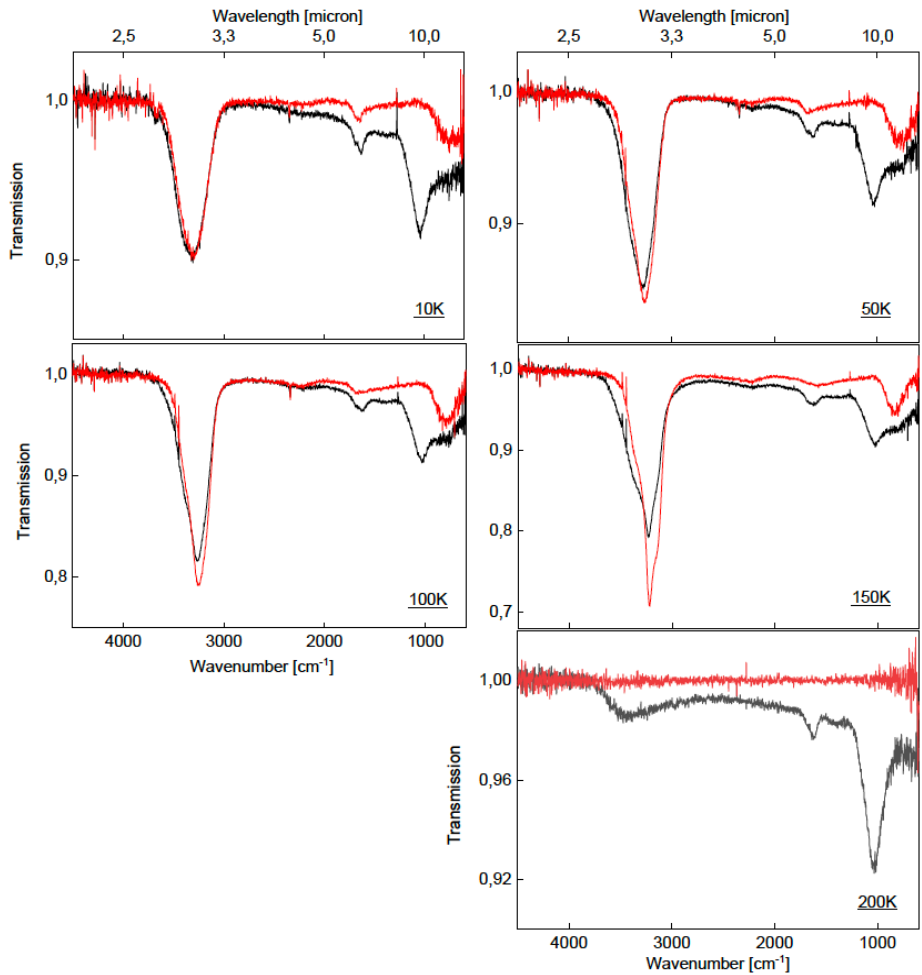

Figure 1. The IR spectra of the $H_2O$ (red curves) and the $MgSiO_3/H_2O$ (black curves) samples at 10, 50, 100, 150, and 200 K.

The most interesting cases are 150 K, after the transformation of the water ice structure from amorphous to crystalline and 200 K. 200 K is above the desorption temperature of water ice (160 – 180 K depending on the ice thickness and the heating rate). However, the $H_2O$ signatures are clearly visible in the spectra of the silicate/ice mixtures. Such trapping of water molecules on silicate grains was already shown in our previous studies [10,11]. From our point of view, trapped water cannot be considered as water ice and presents, probably, water molecules strongly bound in hydrophilic binding sites on the silicate surface. It is known that OH-stretching in hydrated silicates is blue-shifted with respect to pure $H_2O$-ice [28,29]. Spectra of hydrated silicates show a broad absorption from 2.75–3.2 μm caused by bound water molecules. This region is the region where we observe the water stretching band at 200 K. The observed high-frequency (short-wavelength) broadening of the 3.1-micron band at 50, 100, and 150 K can be also explained by the presence of water molecules strongly bound on the silicate surface. Experimental evidence of the trapping of water molecules on silicates beyond the ice desorption temperature reinforces the result of the calculations on the interaction of water with silicates, which demonstrated the presence of stronger adsorption sites on the surface, where water molecules may be retained inside the snow line [30,31].

Extended Data Figure 1 presents the 150 K spectra of three silicate/$H_2O$ samples and of the $MgSiO_3$/$H_2O$ samples with the mass ratios of 2.7 and 7.5. For the silicate/$H_2O$ samples, we observe only slight differences in the spectra mainly related to the shift of the Si-O stretching band and the baseline. For the spectra of the 2.7 and 7.5 $MgSiO_3$/$H_2O$ samples, the observed differences are mainly related to the baseline. All these differences do not influence noticeably the fit of the laboratory spectra to the observational ones in the 3.1 μm region, on which the following discussion will be focused. For this reason, we concentrate further on the $MgSiO_3$/$H_2O$ 7.5 mixture as a qualitative representative of all silicate/ice mixtures studied. A quantitative study of the influence of different types of silicates as well as of the silicate/ice mass ratio on the spectral properties of silicate/ice mixtures is beyond the scope of the present work.

An approximately linear $MgSiO_3$/$H_2O$ mass ratio dependence of the relative (to deposited) remaining amount of water in the silicate-grains/water-ice samples at 200 K was obtained previously in the range of the mass ratios 0.3 – 2.7 [11]. In the present study, we extended the mass ratio range up to 10.5. As the reader can see in Extended Data Figure 2, the dependence stays linear independently on the type of silicate grains. The value for the relative remaining amount of water reaches, for the highest mass ratio, 22%. This corresponds to the silicate/$H_2O$ mass ratio of about 50. Additional measurements at 250 and 300 K have shown that the amount of trapped water decreases with the temperature increase by a factor of about 1.5 for 250 K and of about 4 for 300 K as compared to 200 K.

**Comparison to observations of protostellar envelopes**

As the strongest spectral variations can be seen in the 3.1 μm water ice band, and this band provides the cleanest measure of the $H_2O$ ice column density in astrophysical environments, we will focus our comparisons with astronomical observations on this particular band. For this comparison, we selected three lines of sight toward the protostars Orion BN, Orion IRc2, and Mon R2 IRS3, observed with the Infrared Space Observatory (ISO) [32], and the protoplanetary disk d216-0939 observed with the Subaru Telescope [33]. We selected the sources because they have high signal-to-noise spectra and show evidence for crystalline water ice in their spectra. This latter criterion is because crystalline water ice forms at a high temperature (150 K on the laboratory timescale) close to the evaporation temperature and that at this temperature the clearest differences can be observed between the pure ice and silicate/ice spectra (see Figure 1). The spectra of the four sources as well as the comparisons to our laboratory measurements are presented in Figures 2 (the protostars) and 3 (the disk). In the case of protostars, first, we fitted a low-order polynomial to the observed continuum emission outside the ice absorption band (the top panels in Figure 2) and subtracted the fitted continuum of the spectra and converted the continuum subtracted values to apparent optical depth (middle and lower panels in Figure 2). In the case of the disk, the normalized optical depth data were used. Next, we compared the observed optical depth to our 150 K spectra. As one can see from the middle panels in Figure 2 and the upper panel in Figure 3, the $MgSiO_3/H_2O$ measurements show a much broader profile than the pure ice measurements and can explain a substantial part of the blue wing of the observed absorption bands.

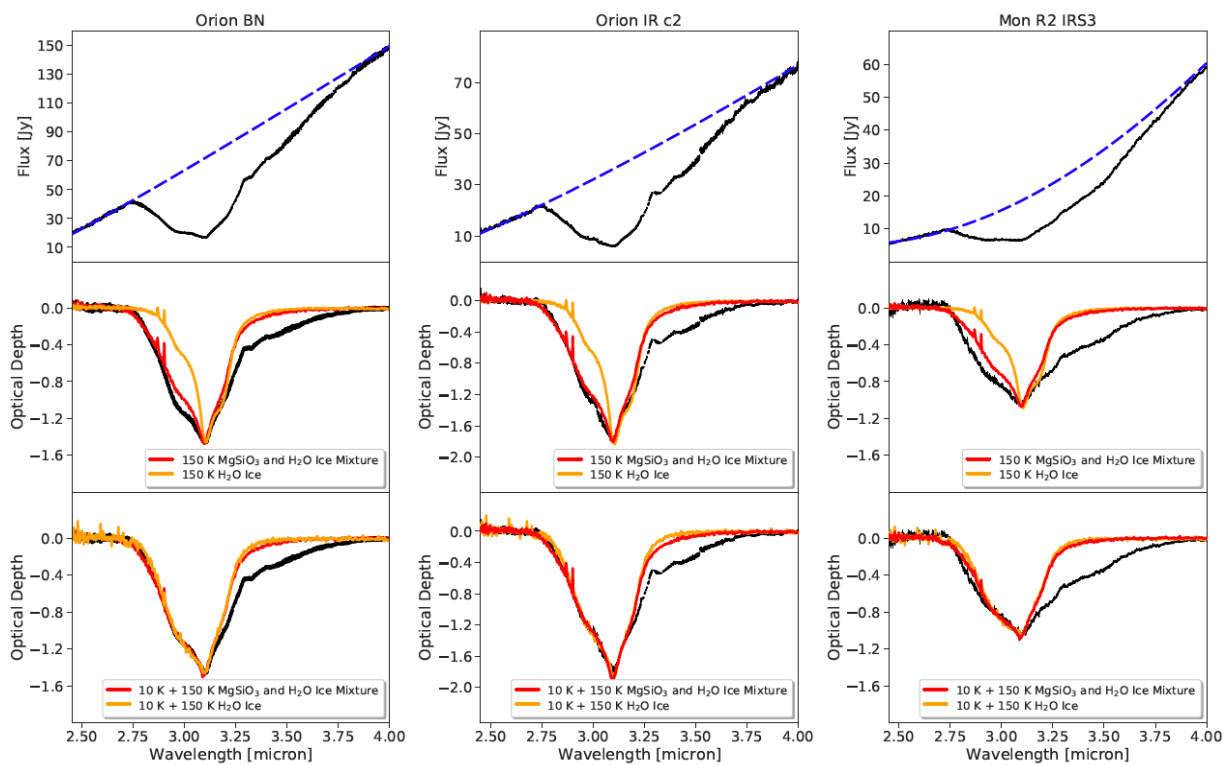

Figure 2. Comparisons between our laboratory measurements and the observational spectra of protostars. Upper panels: the ISO observational spectra of Orion BN, Orion IRc2, and Mon R2 IRS3 [32] with estimated continua. Middle panels: the normalized laboratory spectra of the $H_2O$ (orange) and $MgSiO_3/H_2O$ (red) samples at 150 K compared to the observational spectra. Lower panels: the fits of the observational spectra by a mathematical mixing of the $H_2O$ (orange) and $MgSiO_3/H_2O$ (red) 150 and 10 K spectra. The mixing ratios are presented in Table 1.

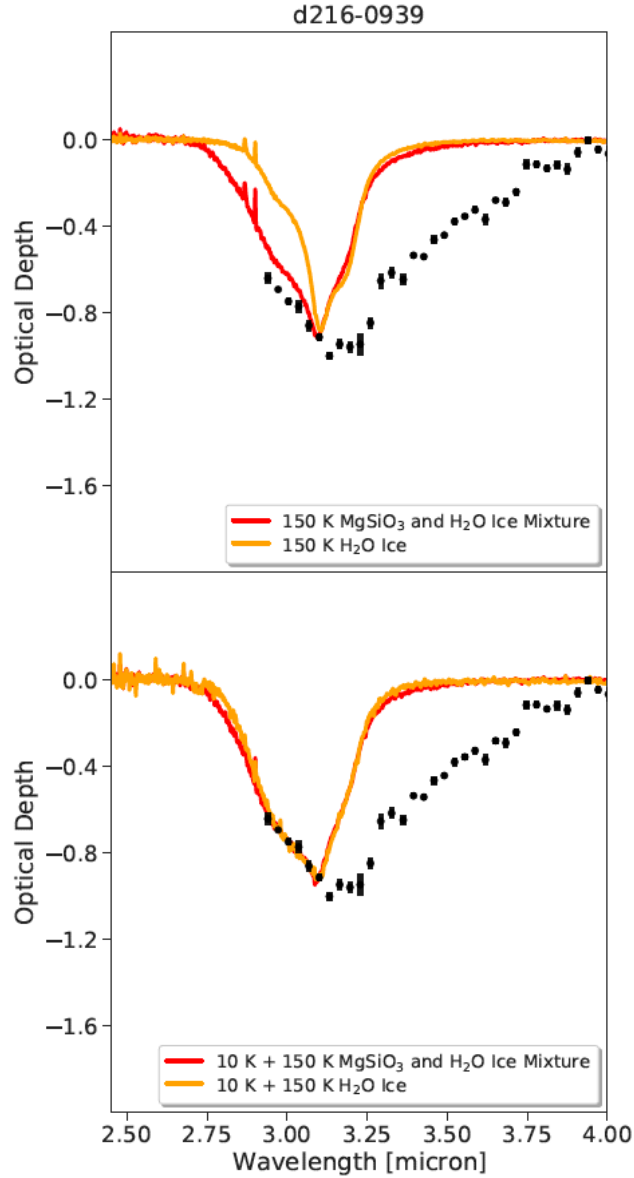

Figure 3. Comparisons between our laboratory measurements and the observational spectrum of d216-0939. Upper panel: the normalized laboratory spectra of the $H_2O$ (orange) and $MgSiO_3/H_2O$ (red) samples at 150 K compared to the observational spectrum of d216-0939. Lower panel: the fits of the observational spectra by a mathematical mixing of the $H_2O$ (orange) and $MgSiO_3/H_2O$ (red) 150 and 10 K spectra. The mixing ratios are presented in Table 1.

The origin of the red wing of the 3.1-micron water ice band is a long-standing question. The formation of ammonia hydrates in combination with scattering on large, micron-sized, particles, may partly explain the red wing of the observational water ice band (for review see [34]). In addition, the OH-stretching band of water ice produced by $O_2$–H co-deposition has a more intense low-frequency shoulder compared to $H_2O$ ice produced by vapour deposition [35] and may also contribute to the red wing. We note that mixing with silicates explains the blue- but not the red wing of the band.

As the silicate and ice grains in protostellar envelopes and protoplanetary disks will most likely have a broad temperature range depending on their proximity to the star, we also fitted the observational spectra by a mathematical mixing of the laboratory 150 and 10 K spectra to mimic such a temperature distribution of the absorbing material. Thus, we applied the simplest multi temperature model possible, namely a two-temperature model consisting of cold (amorphous ice) and warm (crystalline ice) material aiming at making a quantitative comparison between two different grain models: pure ice and ice physically mixed with silicates. The true temperature distributions are, of course, more complex.

The results are presented in the lower panels of Figures 2 and 3 and in Table 1 and show immediately that using pure ice opacities requires a substantial amount of cold ice to explain the blue wing of the 3-micron complex, in contrast to the mixed grain model. The mixing ratios for the 150:10 K spectra and derived mass average temperatures are presented in Table 1 together with formal errors to the quantities derived from the least square fit. However, the reader has to realize that given the very simple nature of the model we use, the true mass temperature ratios will be different and these values should not be taken at face values.

Table 1. The mixing ratios M for the $H_2O$ and $MgSiO_3/H_2O$ 150:10 K spectra and derived mass average temperatures of the protostellar envelopes and protoplanetary disk.

| Source | $H_2O$ | | $MgSiO_3 + H_2O$ | |
|---|---|---|---|---|
| | M 150K/(150K+10K) | T (K) | M 150K/(150K+10K) | T (K) |
| Orion BN | 0.165 ± 0.002 | 33.2 ± 0.2 | 0.642 ± 0.006 | 99.9 ± 0.8 |
| Orion IRc2 | 0.236 ± 0.003 | 43.1 ± 0.4 | 1.000 ± 0.016 | 150.0 ± 2.3 |
| Mon R2 IRS3 | 0.133 ± 0.004 | 28.6 ± 0.6 | 0.520 ± 0.011 | 82.8 ± 1.5 |
| d216-0939 | 0.161 ± 0.010 | 32.6 ± 1.4 | 0.565 ± 0.020 | 89.1 ± 2.8 |

From this comparison, it can be clearly seen that the blue wing of the 3.1-micron ice absorption band can be equally well fitted by mixing the 150:10 K spectra of the $H_2O$ or $MgSiO_3/H_2O$ samples. However, the mixing ratios in these two cases differ substantially (see Table 1). Where the fit by mixing the pure $H_2O$ spectra would lead to a conclusion that mainly the low-temperature ice (more than 80%) is present in the protostellar envelopes and protostellar disk, by mixing the silicate/ice spectra the low-temperature ice accounts only for about 30 to 50% of the column mass toward the investigated objects. In the case of Orion IRc2, the best fit was obtained by using only the $MgSiO_3/H_2O$ 150 K spectrum (0% of the low-temperature ice). The derived mass average temperatures of the envelope and disk materials in the line of sight differ by a factor of three, being between 28 and 43 K when using pure water ice and between 83 and 150 K when using the $MgSiO_3/H_2O$ data. The higher values are reasonable for the protostellar envelopes around the three sources investigated and for the protoplanetary disk and show that there is a strong dependence of the derived physical quantities of the absorbing material on the grain model used. The higher temperature values are also supported by the absence of the CO ice feature in the spectra of the envelopes [32] as CO ice desorbs around 40 K. If the observed ice and silicate dust in these environments is better represented by a mixture, the conclusions of absorption studies toward these classes of objects will need to be substantially revised.

**Comparison to observations of the diffuse ISM**

Apart from the protostellar and protoplanetary sources, we also made comparisons to the observations along the sightline towards Cygnus OB2 12, probing the diffuse interstellar medium, done by ISO [32] and Spitzer [36]. The results are shown in Figure 4. The top panel shows the comparison between the apparent optical depth derived from ISO (shorter than 5 microns) and Spitzer (longer than 5.5 microns) observations and our laboratory measurements of the mixture of solid-state water and silicates at 200 K. Such a mixture should be the closest analogue to dust grains in the diffuse ISM, not because of the temperature but because of a possible presence of strongly bound water molecules and absence of volatile species.

The presence of water on dust grains in diffuse and translucent clouds was predicted by a number of models (e.g., [37,38]). While absorption bands in the 3-micron region have been detected in the diffuse ISM, those have been attributed to amorphous, hydrogenated carbonaceous dust (e.g. [39]) and not to water ice. In an earlier study [40], a shallow feature near 2.75 μm was claimed, indicating, according to the authors, the presence of hydrated silicates (phyllosilicates). However, this result was not confirmed in Gibb et al. [32]. The detection of broad, shallow features is very sensitive to both spectrophotometric calibration and how one fits the continuum. We opted to fit a low-order polynomial outside the regions of interest over a wider wavelength range. In this case, a shallow, broad absorption feature can be seen between 2.8 and 3.8 microns. In the 200 K laboratory spectra, the $H_2O$ stretching band is blue-shifted with respect to the low-temperature ice band and centred at about 2.9 μm. The observed feature in the Cygnus OB2 12 spectrum agrees well with the laboratory spectrum.

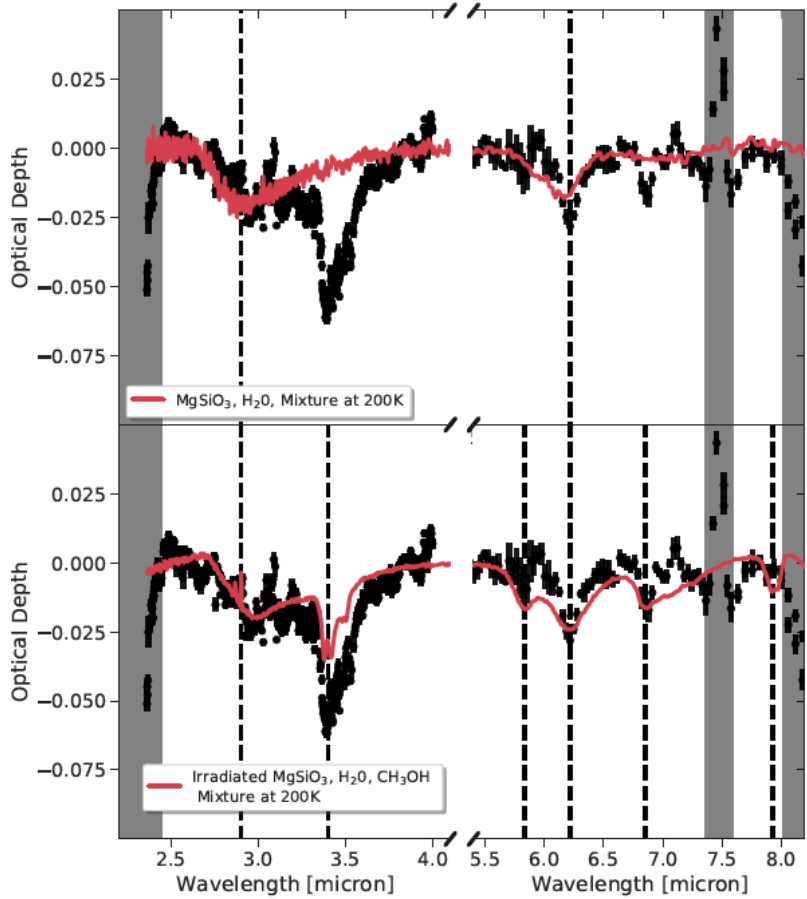

Figure 4. Comparisons between our laboratory measurements and the derived apparent optical depth towards Cyg OB 2 12. The top panel shows the comparison between the apparent optical depth (black curve) derived from ISO (shorter than 5 microns) and Spitzer (longer than 5.5 microns) observations and our laboratory measurements of the mixture of solid-state water and silicates at 200 K (red curve). The lower panel shows the comparison to the laboratory measurements of the UV irradiated mixture of solid-state $H_2O$, $CH_3OH$ and silicates warmed up to 200K. The vertical dashed lines in this figure indicate the observed absorption features in optical depth profiles. The grey area indicates the spectra region dominated by a blend of strong HI emission lines. All plotted error bars represent a 1-sigma uncertainty limit.

However, the main result of the comparison is the matching between the 6.2-micron water band in the laboratory 200 K spectrum and the band in the Spitzer spectrum. This result is the first evidence of the presence of solid-state water in the diffuse medium, where water molecules could be embedded in silicates. The $H_2O$ + $MgSiO_3$ laboratory spectrum in the top panel of Figure 4 corresponds to the $H_2O$ column density $N(H_2O)$ of $3\times10^{16}$ cm$^{-2}$. This value is in a perfect agreement with the upper limit for the column density of water ice of $4\times10^{16}$ molecule cm$^{-2}$ provided by Whittet et al. [40]. Taking $N(H) \approx 4\times10^{21}$ cm$^{-2}$ [41], we find $N(H_2O)/ N(H) \approx 10$ ppm of oxygen "missed" in the diffuse ISM, which corresponds to ~2% of the cosmic O-budget, ~7% of the O in dust, and ~5% of the missing oxygen in the dense ISM [42]. We have to note that our silicate/water mixtures at 200 K are not analogues of phyllosilicates where OH-groups or $H_2O$ molecules are chemically incorporated into crystalline silicates corresponding to a high-temperature treatment. This points out that "water reservoirs" on silicates can be formed in interstellar and circumstellar media at low temperatures.

To fit the other bands in the observational spectrum, we performed a number of experiments aimed at the formation of hydrocarbons on silicates. $MgSiO_3+H_2O+CH_3OH$, $MgSiO_3+H_2O+CO_2$, and $MgSiO_3+H_2O+CH_4$ mixtures (containing C- and H-bearing ice molecules abundant in astrophysical environments [34]) were irradiated by UV at low temperatures and heated up to 200 K to desorb volatile species. All residue spectra show similar profiles containing bands that can be attributed to vibration modes of different organic materials. The resulting substance is a refractory kerogen-like material comparable to an oxygen-containing hydrogenated carbon one. A quantitative study of the influence of different types of parent molecules and the silicate/ice mass ratio on the spectral properties of the residue is beyond the scope of the present work. In the lower panel of Figure 4, we present the result of the best fit.

Addition of a hydrocarbon layer leads to an appearance of new bands and a slight shift of the 6.2-micron band, which, in this case, matches the band at 6.22 in the observational spectrum. The shift may point to the contribution of the 6.25-micron hydrocarbon band, however, this band should be accompanied by other absorption bands at 6.65 and 11.3 microns [39], both are not observed (see also [36]). The other bands coinciding in the laboratory and Spitzer spectra are: around 3.4 μm corresponding to the symmetric and antisymmetric C-H stretching vibrations, 5.83 μm corresponding to C=O stretching, and 6.85 μm corresponding to CH-bending modes in $CH_2$ and $CH_3$ groups (for assignments of the IR bands see [26,43]). One more band existing in the laboratory spectrum and absent in the observational spectrum is 7.95 μm related to C-C stretching and C-H deformation.

A recently submitted paper by Hensley & Draine [44] also re-analyzed the Spitzer and ISO data of Cyg OB2 12. Hensley & Draine base their analysis on the published Spitzer low-resolution spectra of Ardila et al. [45] and one of the ISO Spectra published by Sloan et al. [46] (TDT 13901048). They claim the detection of a number of carbonaceous features at 3.3, 3.4, 6.2, 6.85, and 7.7 microns. In our new reduction of the Spitzer observations, we see no evidence for an absorption band at 7.7 microns. We can confirm the presence of the absorption bands at 6.2 and 6.85 microns, however, we assign the 6.2 band mainly to solid-state water with a possible slight contribution from carbon. A weak 3.3-micron band could indeed be present. In addition, we detected a band at 5.83 microns.

**Discussion**

One can think of multiple formation and processing scenarios leading to different grain structures, like bare silicate and ice grains or a layer surrounding a compact silicate core (both mean physical separation of ice and silicate dust) or ice/dust mixtures. To prove which scenario is the correct one or if multiple scenarios exist, these different models need to be compared to the observations, for which the opacities of the solid-state materials need to be known. In this paper we study the grain material of well mixed ice and silicates, a structure, which could occur if a condensation of both materials takes place at low temperatures. This might occur after shocks traveling through cold environments or when submicron grains containing both ice and silicates grow into larger grains.

While our laboratory measurements of ice/dust mixtures at 10 K show no difference in the 3-micron spectral region compared to pure ice measurements - in this case, it is not possible to distinguish between different grain models - at higher temperatures significant differences occur, leading to substantial differences in mass-temperature estimates. Our laboratory data can explain the observations giving reasonable mass average temperatures for the protostellar envelopes and protoplanetary disk considered and reinforce previous studies discussing the dust/ice mixing in cold astrophysical environments. The agreement between the laboratory silicate/ice spectra and the observational spectra demonstrates that a substantial fraction of solid-state water may be mixed with silicate grains in the diffuse ISM, protostellar envelopes, and protoplanetary disks.

The large differences in derived temperature distributions for the different grain models is, in our opinion, a very important message to the wider astronomical community that only by determining the exact structure of the solid-state material full results can be obtained. However, the exact nature (including structure and composition) of the solid-state materials in different astrophysical environments can only be determined by spatially resolved multi-wavelength observations over a wide wavelength range and by modelling both the thermal emission as well as the absorption towards the central object. New facilities like JWST are ideally suited to provide us with such data. To fully use these observation, laboratory measurements like those presented in this paper are urgently needed.

According to laboratory experiments, water ice thermally desorbs completely at 160 – 180 K. However, our results show that water mixed with silicates is partly trapped and can stay on silicates at higher temperatures. This result has an implication for the amount of water in grains in protostellar envelopes and protoplanetary disks inside the snow line. Trapped water can survive the transition from cold star-forming regions to planet-forming disks and stay in silicates in the terrestrial planet zone. This reinforces the wet scenario of the origin of water on Earth and terrestrial planets meaning that local planetesimals in the time of the Earth formation retained some water at high temperatures through its physisorption or chemisorption on silicate grains [47].

Though there were predictions and tentative detections of solid-state water, there was no solid evidence that it exists in the diffuse ISM. Our laboratory experiments demonstrate that water can be present on dust grain surfaces in the diffuse ISM and show up as the weak OH-stretching and OH-bending bands around 3 and 6 microns as is demonstrated in Figure 4.

The presence of water on silicates in the diffuse ISM has three important consequences. First is the long-standing problem of the depletion of elemental oxygen [48,49]. As much as a third of the total elemental O budget is unaccounted for in any observed form at the transition between diffuse and dense ISM phases. Solid-state water on silicates in the diffuse ISM is a reservoir of unaccounted oxygen. Thus, together with water, a part of "missed" oxygen can be found.

Second is the question of the formation of dust in the diffuse ISM. In the ISM, dust grains can be completely destroyed by supernova shocks. Estimations showed that only a few percent of the dust produced by stars survive [50]. In this case, the mechanism of the dust grain formation is a cold condensation at interstellar conditions. According to coagulation models [51,52] and collisional experiments [53], ice-coated grains may stick together much more efficiently. Thus, a detection of water and an estimation of its abundance is important for understanding of the mechanism and efficiency of grain growth.

Third is that the detection of $H_2O$ shows a possible pathway to the formation of simple and complex molecules on silicate grains mixed with carbon. This idea is based on the laboratory experiments, such as the formation of CO and $CO_2$ involving atoms of different carbon surfaces covered by $H_2O$ ice by UV, ion, and proton irradiation and the formation of formaldehyde $H_2CO$ by the addition of O and H atoms to bare carbon grains (see [5] and references therein). If solid-state water exists in the diffuse ISM, the mentioned experiments should present reliable pathways for the formation of molecules in surface reactions in interstellar clouds. These pathways may explain the detections of COMs, such as $CH_3OH$ [54], $HC_5N$ and $CH_3CN$ [55] in diffuse and translucent clouds.

Our laboratory measurements from 2 to 25 μm cover the major parts of the spectral ranges of the NIRSPEC (0.6 – 5.5 μm) and MIRI (5 – 28 μm) instruments of JWST presenting reliable data for the decoding and analysis of future JWST spectra. We hope that JWST will be able, with the help of the new laboratory data, to bring more information on solid-state water where we do not expect it, in the diffuse medium and in warm/hot regions of protostellar envelopes and planet-forming disks inside the snowline.

Corresponding author: Alexey Potapov, *alexey.potapov@uni-jena.de*

**Acknowledgments**

We are grateful to Erika Gibb and Hiroshi Terada for providing us the observational data and to three anonymous reviewers for questions, suggestions, and corrections that helped to improve the manuscript. This work was supported by the Research Unit FOR 2285 "Debris Disks in Planetary Systems" of the Deutsche Forschungsgemeinschaft (grant JA 2107/3-2). TH acknowledges support from the European Research Council under the Horizon 2020 Framework Program via the ERC Advanced Grant Origins 83 24 28.

**Author contributions**

A.P. led the project. A.P. and J.B. designed the research and wrote the paper. A.P. performed the laboratory experiments. J.B. performed the analysis of the observational data. C.J. and T.H. participated in data interpretation and discussion.

**Competing interests**

The authors declare no conflict of interest.

**Methods**

**Laboratory experiments**

Our experimental set-up and procedure have been presented in two recent papers devoted to dust/ice mixtures [10,11]. In the present study, nanometre-sized enstatite $MgSiO_3$, forsterite $Mg_2SiO_4$, and olivine $MgFeSiO_4$ amorphous silicate grains were produced by laser ablation of Mg:Si 1:1, Mg:Si 2:1, and Mg:Fe:Si 1:1:1 targets, correspondingly, in quenching atmospheres of 4 Torr $O_2$ for $MgSiO_3$ and $Mg_2SiO_4$ grains and He+$O_2$ in the volume ratio of 3:1 for $MgFeSiO_4$ grains. After the formation, grains were extracted from the ablation chamber and deposited, simultaneously with water vapours, on KBr substrates at the temperature of 10 K and the pressure of $5\times10^{-8}$ mbar in the deposition chamber. Grains interact on the substrate forming a porous layer of fractal aggregates with the sizes of up to several tens of nm.

The thickness of the grain deposits was controlled by a quartz crystal resonator microbalance using known values for the deposit area of 1 cm$^2$ and density. The thickness of the water ice was calculated from the 3.1 μm water band area using a band strength of $2\times10^{-16}$ cm molecule$^{-1}$ [56]. The dust/ice mass ratios were 2.7 and 7.5 for $MgSiO_3/H_2O$, 4.5 and 9.0 for $MgFeSiO_4/H_2O$, and 10.5 for $Mg_2SiO_4/H_2O$, and were calculated from the thicknesses of the deposits using the densities of 1.1 g cm$^{-3}$ for high-density amorphous water ice [57], and 2.5 g cm$^{-3}$, 2.9 g cm$^{-3}$, and 3.7 g cm$^{-3}$ for $MgSiO_3$, $Mg_2SiO_4$, and $MgFeSiO_4$ silicates correspondingly. One more sample was prepared to mimic a layer of hydrocarbons on cosmic silicates. The mixture $MgSiO_3/CH_3OH/H_2O$ with the mass ratio of 4 for $MgSiO_3/CH_3OH$ and 10 for $CH_3OH/H_2O$ was deposited at 10 K and irradiated for 2 hours by a broadband deuterium lamp (L11798, Hamamatsu) with a flux of $10^{15}$ photons s$^{-1}$ cm$^{-2}$. The lamp has a broad spectrum from 400 to 118 nm with the main peaks at 160 nm (7.7 eV) and at about 122 nm (10.2 eV) (with the intensity of about 20% of the 160 nm peak).

After the deposition, substrates were warmed up consequently to 50, 100, 150, and 200 K. The samples were allowed to stabilize at each temperature for 30 minutes. IR spectra were taken in-situ using an FTIR spectrometer (Vertex 80v, Bruker) in the transmission mode. The spectra of KBr substrates recorded before the depositions at 10 K were used as reference spectra.

**Observational data**

For the comparison of our laboratory spectra to observational spectra of protostellar envelopes, we selected three line of sights toward the protostars Orion BN, Orion IRc2, and Mon R2 IRS3 with published high quality Infrared Space Observatory (ISO) data. In our analysis we used these spectra as published, so for detailed descriptions of these three sources as well as details on the data reduction and calibration we refer to the Gibb et al. [32] study. For the comparison between our laboratory measurements and this data set, we followed a similar approach as [32]. First, we fitted a low-order polynomial to the observed continuum emission outside the ice absorption band and divided the observed spectra with the fitted continuum to derive an apparent optical depth. Note, that we applied a sigma clip to remove any gas lines which might influence our continuum determinations. At wavelengths where this was not possible (short ward of 2.5 micron due to blending) we decided not to use these wavelengths in or analysis or fits. Next, we compared the derived optical depth to our laboratory transmission spectra by converting these latter measurements to absorption coefficients, assuming that the reflection in these measurements was zero.

For the comparison to observations of the diffuse ISM, we selected mid-infrared spectroscopic observations by ISO towards Cygnus OB2 12. Gibb et al. [32] also published observation of this target, however, they only used one out of three ISO observations of this source covering the 3 μm ice band. Given the relatively weak absorption bands towards this source, we opted to use the ISO spectra published by Sloan et al. [46]. The three observations (TDT 33504130, 03602226, and 13901048) can be seen in Extended Data Figure 3. For details on the data reduction, we refer to Sloan et al. [46]. We used their spectra as published, with small corrections to the absolute flux calibration to match the spectroscopic observations of this target with the low-resolution spectrograph onboard the Spitzer Space Telescope. We corrected the absolute flux values by 0.94, 0.99 and 0.985 for TDT 33504130, 03602226 and 13901048, respectively. We also noticed that spectra published by Sloan et al. had a small jump between the short wavelength and longer wavelength bands (see [46] Table 1 for the band definitions). We corrected this by shifting band 1A to 1E by a factor of 1.03 for TDT 33504130 and TDT 13901048. For TDT 03602226 we corrected the fluxes in bands 1A and 1B by a factor of 1.03. We then fitted a low order polynomial to the ISO-SWS spectra short ward of 8 microns, not using those wavelengths where absorption bands are expected. The fitted polynomial continua for the three observations are also plotted in Figure 4. Using the fitted continua, we then derived the apparent optical depth following the same procedure as with the other sources. We combined the optical depth profiles of the three observations and re-binned the spectra to a uniform wavelength grid, taking into account the errors on the individual data points, to gain signal-to-noise. The final optical depth profile is shown in the lower panel of Extended Data Figure 3.

The highest SNR is reached for wavelengths shorter than 4.08 microns, (band 1A to 1E of SWS) which use an InSb detector (see also [46]) in contrast to the longer wavelength data which was observed using a Si:Ga detector. We therefore only use the ISO data shorter than 4.08 micron in this study. For the longer wavelengths shown here, we will use the Spitzer-IRS observations which have a superior signal-to-noise.

Cyg OB2 12 was observed with the Infrared Spectrograph (IRS) [58] onboard the Spitzer Space Telescope at 2 epochs; 6 August 2008 (AOR KEY: 27570176, PID: 485) and 10 October 2004 (AOR KEY: 9834496, PID: 179), using the low (R~60-120) as well as the high (R~600) resolution spectroscopic mode. These Spitzer data were used by [36] for the analysis of the observed silicate absorption bands through the diffuse ISM towards Cyg OB2 12. We refer to this paper for details on the Spitzer observations. As the focus of this paper is not the strong silicate absorption bands but the much weaker bands of solid-state water and carbonaceous materials, we decided to re-calibrate the observations using the data reduction packages developed for the c2d (core to discs) and feps (formation and evolution of planetary systems) Spitzer key projects (see [59,60] for details and application of the data reduction packages). In brief, we applied the following calibration steps to the low and high-resolution data:

All data products were processed through the Spitzer Science Centre (SSC) pipeline version S18.18.0. In the case of low-resolution mode, the data reduction process started from the *droopres* intermediate data product. For the high-resolution data we used the *rsc* data product as a starting point. The background emission in the low-resolution data has been subtracted using associated pairs of imaged spectra from the two nod positions, also eliminating stray light contamination and anomalous dark currents. Pixels flagged by the data pipeline as being "bad" were replaced with a value interpolated from an in the dispersion direction elongated 8~pixel perimeter surrounding the flagged pixel. The spectra were extracted using a 6~pixel and 5~pixel fixed-width aperture in the spatial dimension for the Short Low and the Long Low modules, respectively. The low-level fringing at wavelengths >20 μm was removed using the irs fringe package. The spectra were calibrated with a position-dependent spectral response function derived from IRS spectra and MARCS stellar models for a suite of calibrators provided by the Spitzer Science Centre. To remove any effect of pointing offsets in the short-low module data, we matched orders based on the point spread function of the IRS instrument, correcting for possible flux losses.

For the high-resolution data, we applied an optimal source profile extraction method that fits an analytical PSF derived from sky-corrected calibrator data and an extended emission component, derived from the cross-dispersion profiles of the flat-field images, to the cross-dispersed source profile. It is not possible to correct for the sky contribution in the high-resolution spectra, subtracting the two nod positions as with the low-resolution observations, due to the small slit length. We used the background estimate from the source profile fitting extraction method to remove the background emission. For correcting "bad" pixels we used the IRSCLEAN package. We further removed low-level (~1%) fringing using the irs fringe package. The flux calibration for the high-resolution spectrograph has been done in a similar way as for the low-resolution observations. For the relative spectral response function, we also used MARCS stellar models and calibrator stars provided through the SSC. The spectra of the calibration stars were extracted in an identical way to our science observations using both extraction methods. As with the low-resolution observations, we also corrected for possible flux losses due to pointing offsets. We estimate the accuracy of our flux estimates to be at the level of a few %, consistent with the uncertainties on the used stellar models of the calibration stars.

After these steps, we noticed a 2% flux difference between the short wavelength modules and the long-wavelength modules of the low-resolution spectrograph which we corrected by shifting the short-wavelength spectra upwards by this amount. The high-resolution data matched the low-resolution long-wavelength spectra perfectly. A possible cause for the jump could be the difference in slit width between the short and long wavelength modules and slightly extended target or non-uniform background. We also noticed a slight mismatch of 1.5% between the second- and first-order spectra of the low-resolution spectrograph. We corrected this by shifting the second-order upwards by this amount relative to the first-order spectra. A possible reason why there was still a small flux difference between the short wavelength orders even after using our order matching algorithm is probably due to the fact that in the overlapping wavelengths a strong blend of HI lines is visible. Given that the spectral orders have different spectral resolutions, this could lead to slightly different flux levels for a given wavelength, something our matching routine does not take into account as it is based on continuum observations. We also noted that for one of the nods positions of the low-resolution spectroscopic data the detector covering the ~9- and 10-micron spectral range saturated during exposures, leading to incorrect flux estimates. We removed the saturated data from the final spectrum.

The resulting spectra of Cyg OB2 12 can be seen in Extended Data Figure 4 where the top panel shows the spectra of the different orders and modules together with the fitted continuum, which we assumed to be a power-law which was matched to the spectra between 5.5 and 8 microns, excluding those regions were a possible absorption band could occur. The lower panel shows the derived optical depth profile.

**Extended Data**

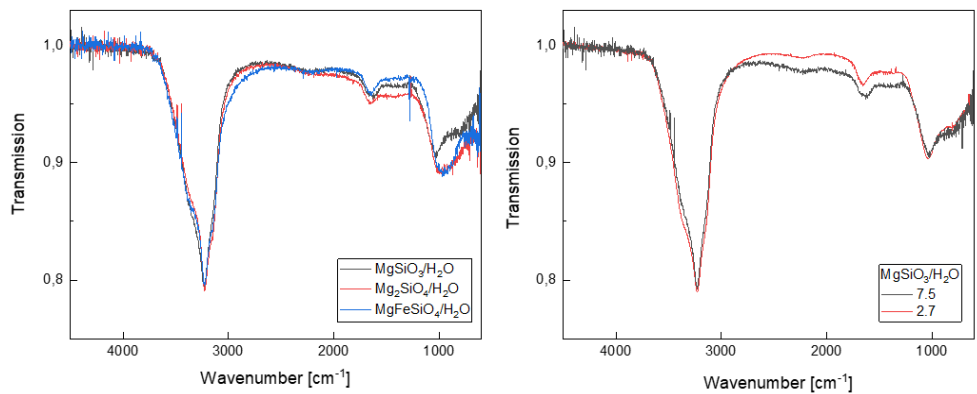

Extended Data Figure 1. The normalized 150 K spectra of the silicate/ice mixtures. $MgSiO_3/H_2O$ 7.5, $Mg_2SiO_4/H_2O$ 10.5, and $MgFeSiO_4/H_2O$ 9.0 samples (left) and the $MgSiO_3/H_2O$ samples with the mass ratios of 2.7 and 7.5 (right).

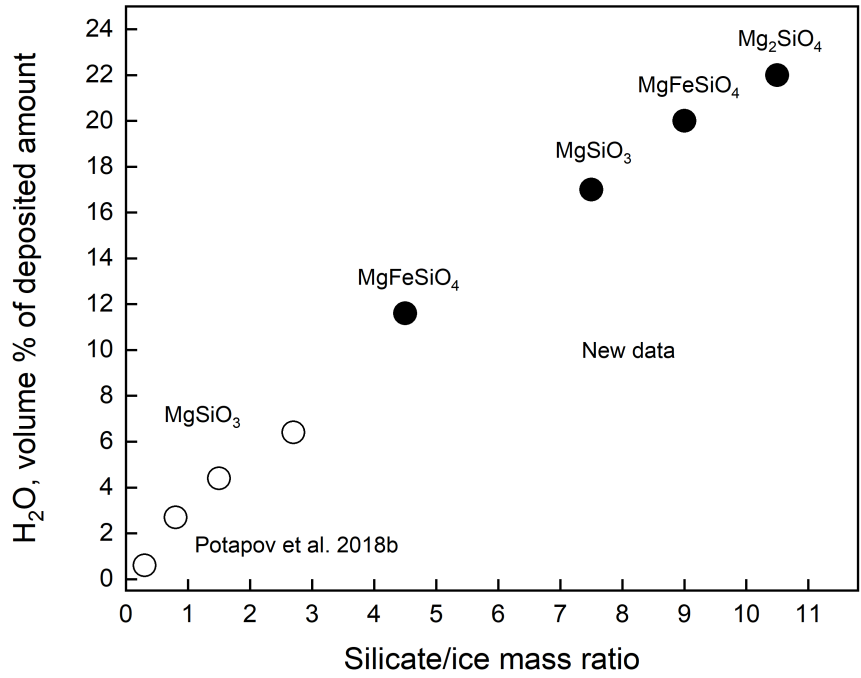

Extended Data Figure 2. Amount of trapped water. Dependence of the volume percentage of remaining water in the silicate/H$_2$O samples at 200 K on the silicate/H$_2$O mass ratio.

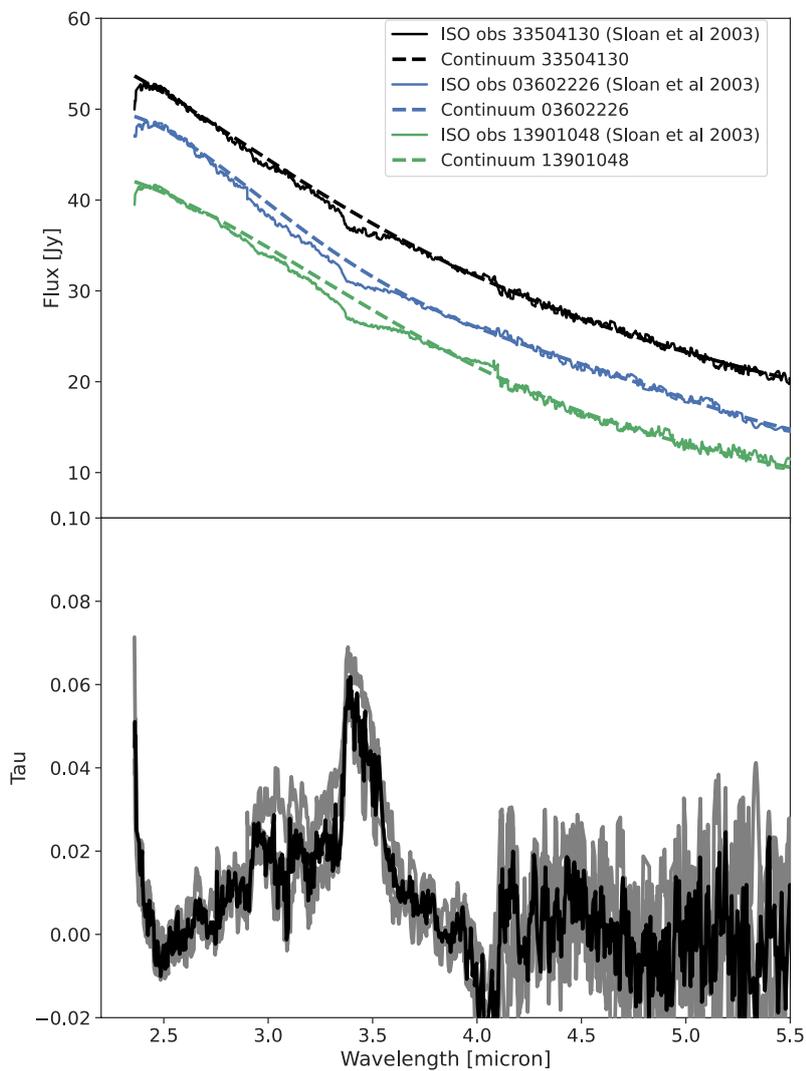

Extended Data Figure 3. ISO-SWS observations of Cyg OB2 12. The top panel shows the three ISO observations of Cyg OB 2 12 (TDT 33504130, 03602226, and 13901048). The two lower spectra are offset by 5, respectively 10 Jy for clarity. Also, show in this panel are the fitted polynomial continua for the three observations. The lower panel shows the derived optical depth for the three observations (gray curves) and the re-binned average (black curve).

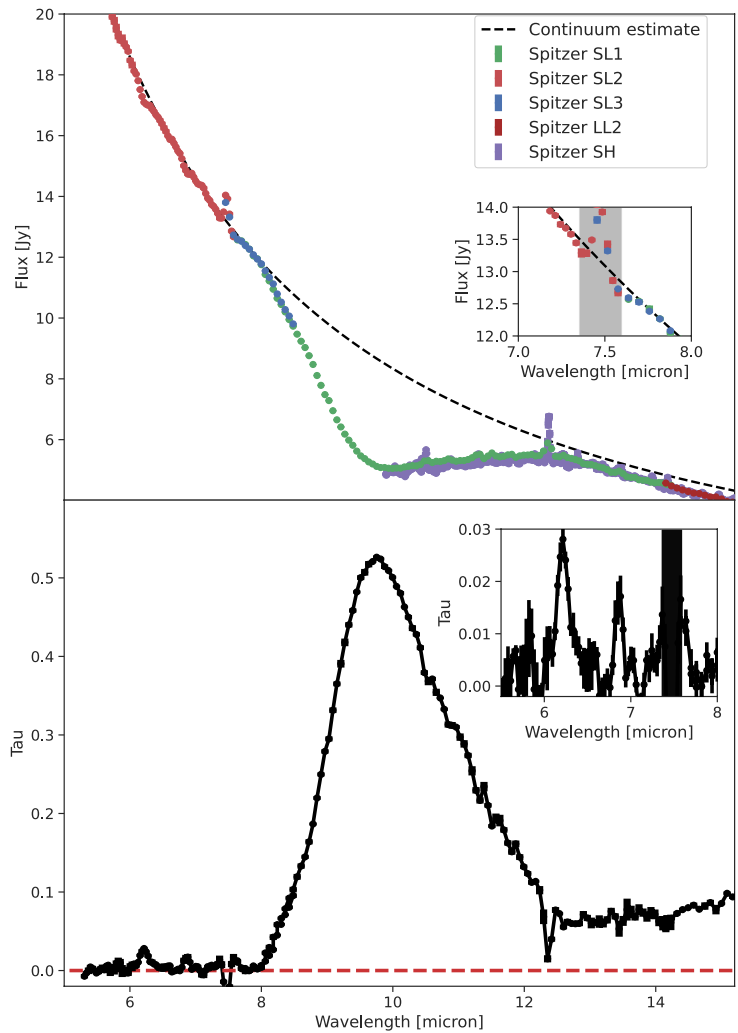

Extended Data Figure 4. Spitzer IRS observations of Cyg OB2 12. The top panel shows the data taken in the 3 spectral orders of the short-wavelength module of the low-resolution spectrograph (green, red and blue data points) the short-wavelength module of the high-resolution spectrograph (magenta data points) and the second order of the long-wavelength module of the low-resolution spectrograph (brown data points). Also plotted in his panel is the fitted power-law continuum (black dashed line). The inset figure shows a zoom of the spectral region between 7 and 8 microns. The grey shaded area indicates the position of a blend of HI emission lines. The lower panel shows the derived apparent optical depth profile. The inset panel shows a zoom of the 5.5 to 8 microns spectral region. All plotted error bars represent a 1-sigma uncertainty limit.

**Data Availability**

The data that support the plots within this paper and other findings of this study are available from the corresponding author upon reasonable request.